\documentclass{aa}  

\usepackage{graphicx}
\usepackage{txfonts}
\usepackage{threeparttable}
\usepackage[colorlinks=true, linkcolor=blue, citecolor=blue, urlcolor=blue]{hyperref}
\begin{document}

   \title{A post-common-envelope binary with double-peaked Balmer emission lines from TMTS}

   \author{Qichun Liu
          \inst{1},
          Xiaofeng Wang\inst{1},
          Jie Lin\inst{2,3,1},
          Chengyuan Wu\inst{4,5,6},
          Chunqian Li\inst{7,8},
          Alexei V. Filippenko\inst{9},
          Thomas G. Brink\inst{9},
          Yi Yang\inst{1,9},
          Weikang Zheng\inst{9},
          Cheng Liu\inst{10},
          Cuiying Song\inst{1},
          Mikhail Kovalev\inst{4},
          Hongwei Ge\inst{4,5,6},
          Fenghui Zhang\inst{4,5,6},
          Xiaobin Zhang\inst{11},
          Qiqi Xia\inst{1},
          Haowei Peng\inst{1},
          Gaobo Xi\inst{1},
          Jun Mo\inst{1},
          Shengyu Yan\inst{1},
          Jianrong Shi\inst{11,12},
          Jiangdan Li\inst{4,5,6},
          \and
          Tuan Yi\inst{13, 14}
          }

   \institute{Physics Department, Tsinghua University, Beijing 100084, People's Republic of China\\
              \email{wang\_xf@mail.tsinghua.edu.cn}
         \and
             CAS Key laboratory for Research in Galaxies and Cosmology, Department of Astronomy, University of Science and Technology of China, Hefei, 230026, People's Republic of China\\
             \email{linjie2019@ustc.edu.cn}
        \and
             School of Astronomy and Space Sciences, University of Science and Technology of China, Hefei, 230026, People's Republic of China
        \and
             Yunnan Observatories, Chinese Academy of Sciences, Kunming 650216, People's Republic of China\\
             \email{wuchengyuan@ynao.ac.cn}
        \and
             Key Laboratory for Structure and Evolution of Celestial Objects, Chinese Academy of Sciences, Kunming 650216, People's Republic of China
        \and
             International Centre of Supernovae, Yunnan Key Laboratory, Kunming 650216, People's Republic of China
        \and 
             School of Physics and Astronomy, Beijing Normal University, Beijing 100875, People's Republic of China
        \and 
             Institute for Frontiers in Astronomy and Astrophysics, Beijing Normal University, Beijing 102206, People's Republic of China
        \and
             Department of Astronomy, University of California, Berkeley, CA 94720-3411, USA
        \and
             Beijing Planetarium, Beijing Academy of Science and Technology, Beijing 100044, People's Republic of China
        \and
             CAS Key Laboratory of Optical Astronomy, National Astronomical Observatories, Chinese Academy of Sciences, Beijing 100101, People's Republic of China
        \and
             School of Astronomy and Space Science, University of Chinese Academy of Sciences, Beijing 100049, People's Republic of China
        \and
             Department of Astronomy, School of Physics, Peking University, Beijing 100871, People's Republic of China
        \and
             Kavli Institute of Astronomy and Astrophysics, Peking University,  Beijing 100871, People's Republic of China        
             }

   \date{Draft: March 10, 2025}
 
  \abstract
   % context heading (mandatory)
   {The dynamical method provides an efficient way to discover post-common-envelope binaries (PCEBs) with faint white dwarfs (WDs), thanks to the development of time-domain survey projects. As close binary systems undergo a common-envelope phase, they offer unique opportunities to study the astrophysical processes associated with binary evolution. }
   {We perform a comprehensive analysis of the PCEB system TMTS J15530469+4457458 (J1553), discovered by the Tsinghua University-Ma Huateng Telescopes for Survey, to explore its physical origin and evolutionary fate.}
   {This system is characterized by double-peaked Balmer emission lines, and we applied a cross-correlation function  to derive its radial velocity (RV) from a series of phase-resolved Keck spectra. The physical parameters of this binary were obtained by fitting the light curves and RV simultaneously. The locations of the Balmer lines were inferred from Doppler tomography, and a MESA simulation was performed to explore the evolution of this system. }
    {Analyses using the cross-correlation function suggest that this system is a single-lined spectroscopic binary and only one star is optically visible. Further analysis through Doppler tomography indicates that J1553 is a detached binary without an accretion disk. Under such a configuration, the simultaneous light-curve and RV fitting reveal that this system contains an unseen WD with mass $M_{\rm A}=0.56\pm 0.09\, M_{\odot}$, and an M4 dwarf with mass $M_{\rm B}=0.37\pm 0.02\,M_{\odot}$ and radius $R_{\rm B}=0.403^{+0.014}_{-0.015}\, R_{\odot}$. The extra prominent Balmer emission lines seen in the spectra can trace the motion of the WD; these lines are likely formed near the WD surface as a result of wind accretion. 
    According to the MESA simulation, J1553 could have evolved from a binary consisting of a 2.0--4.0\,${M}_{\odot}$ zero-age-main-sequence star and an M dwarf with an initial orbital period $P_i\approx 201-476$\,d, and the system has undergone a common-envelope (CE) phase.  
    After about $3.3\times10^6$\,yr, J1553 should evolve into a cataclysmic variable, with a transient state as a supersoft X-ray source at the beginning. J1553 is an excellent system for studying wind accretion, CE ejection physics, and binary evolution theory.}
   {}


\titlerunning{A PCEB with double-peaked Balmer emission lines}
\authorrunning{Liu et al.}
   \maketitle
%
%-------------------------------------------------------------------
\section{Introduction}\label{sec:intro}
More than half of all stars are found in binary systems, so investigating binary interaction is very important to enrich our understanding of stellar evolution \citep{2020RAA....20..161H}. A post-common-envelope binary (PCEB) is a close binary system that contains a compact object such as a white dwarf (WD) and a less massive companion. Since the more massive star inside the initial main-sequence (MS) binary evolves faster, unstable mass transfer toward the companion occurs \citep[e.g.,][]{2021MNRAS.501.1677H} when it becomes a Roche-filling giant. 
The binary enters a common-envelope (CE) phase and continually loses angular momentum caused by the drag, which leads to the formation of a PCEB \citep{1993PASP..105.1373I,2008ASSL..352..233W, 2011A&A...536A..43N}. With magnetic braking induced by stellar winds \citep{2006ApJ...652..636D, 2022MNRAS.517.4916E}, the PCEB will undergo orbital shrinkage and eventually become a cataclysmic variable (CV) as a second episode of mass transfer starts.

Thanks to the contributions of many survey projects, such as the Sloan Digital Sky Survey (SDSS; \citealt{2000AJ....120.1579Y}) and the Large Sky Area Multi-Object Fiber Spectroscopic Telescope (LAMOST; \citealt{2012RAA....12.1197C}), hundreds of PCEBs and candidates have been identified over the past decade \citep{2013MNRAS.433.3398R, 2014MNRAS.445.1331L, 2014A&A...570A.107R}, although the number of systems with detailed parameters is still quite limited \citep{2021ApJ...920...86K}. 
 CE evolution physics has attracted much attention in the study of binary evolution \citep{2022ApJ...933..137G, 2022MNRAS.517.2867H,2024ApJ...961..202G}. For PCEBs with a low-mass secondary, the CE efficiency, $\alpha,$ is suggested to be low by reconstructing the evolution \citep{2010A&A...520A..86Z, 2022MNRAS.513.3587Z, 2023MNRAS.518.3966S}. 
On the other hand, since direct measurements of mass and radius for a single isolated WD are difficult, PCEBs thus provide additional chances to test the theoretical mass-radius relation for WDs, which can be derived through the gravitational redshift observed near the WD surface \citep{2012MNRAS.419..817P, 2017MNRAS.470.4473P, 2019MNRAS.484.2711R}. 

However, most PCEBs and candidates are selected by their optical colors (in the region between the MS stars and WDs) or spectra (composed of the spectrum of a WD and a MS star), since they are confined to the systems that harbor visible WDs. The dynamical method has the ability to discover faint and massive WDs \citep{2024MNRAS.532.1718Q}, which mitigates the bias in the  observed WD sample. Tsinghua University-Ma Huateng Telescopes for Survey (TMTS; \citealt{Zhang+etal+2020+TMTS_performance,2022MNRAS.509.2362L}) began its minute-cadence monitoring of the LAMOST field in 2020, and has recorded numerous flare stars \citep{2023MNRAS.523.2193L}, short-period pulsating stars \citep{lin+etal+2023+BLAP_NatAs}, eclipsing binaries, and other short-period variables \citep{2023MNRAS.523.2172L, 2024NatAs...8..491L,Liu+etal+2024+TMTS_CVs, 2024MNRAS.528.6997G}. We noticed a new PCEB, TMTS J15530469+4457458 (hereafter J1553),
consisting of an invisible WD and an active M-dwarf star, which are referred to as J1553A and J1553B, respectively. 
Although this source was also independently reported as a binary by \citet{2024AJ....168..217D}, here we conduct a more thorough analysis of the observed properties, including the origin of double-peaked Balmer emission lines and the underlying evolutionary history of this system; our analysis is  aided by new Keck spectra and detailed modeling of the binary evolution. In Sect. \ref{sec:lightcurve}
and \ref{sec:floats}, we describe our photometric and spectroscopic observations. Sect. \ref{sec:result} presents the results of radial velocity (RV) curves and light-curve modeling. The Doppler tomography, evolutionary simulation, and relevant discussions are given in Sect. \ref{spec:discussion}. We summarize our results in Sect. \ref{sec:conclusion}.
%--------------------------------------------------------------------
\section{Photometric observations} 
\label{sec:lightcurve}

Tsinghua University-Ma
Huateng Telescopes for Survey observed this source in the $L$ band continuously for 6.94 hr with a cadence of about 1\,min starting on May 4, 2020 ($\alpha = 15^{\rm hr}53^{\rm m}04.69^{\rm s}$, $\delta = +44^\circ57'45.85''$; J2000). The TMTS light curve shows a period of 1.987\,hr from the Lomb–Scargle periodogram \citep[LSP;][]{1976Ap&SS..39..447L, 1982ApJ...263..835S}. The \textit{g}- and \textit{r}-band photometric observations were also collected from the Zwicky Transient Facility (ZTF; \citet{2019PASP..131a8002B, 2019PASP..131a8003M}) Public Data Release 20 (DR20). Because the ZTF observations covered a period of about 5\,yr, their folded light curve can give a more precise estimate of the period. From the LSP of the \textit{r}-band light curve, we obtain $P_{\rm ph} = 0.08378228 \pm 0.00000003$\,d, which corresponds to an orbital period of $P_{\rm orb}= 0.16756456 \pm 0.00000006$\,d for J1533.

The Transiting Exoplanet Survey Satellite (TESS; \citealt{2015JATIS...1a4003R}) also observed J1553, with an ID of TIC~157365951. We queried the observations in Sectors 28, 50, and 51 from the Mikulski Archive for Space Telescopes at the Space Telescope Science Institute, each lasting $\sim 25$\,d. We note that several flares were recorded in both the ZTF and TESS light curves; they were removed visually in our analysis, being typical phenomena in active M dwarfs such as J1553B. 

\section{Spectroscopic observations} \label{sec:floats}

\begin{figure*}
        \includegraphics[width=2.1\columnwidth]{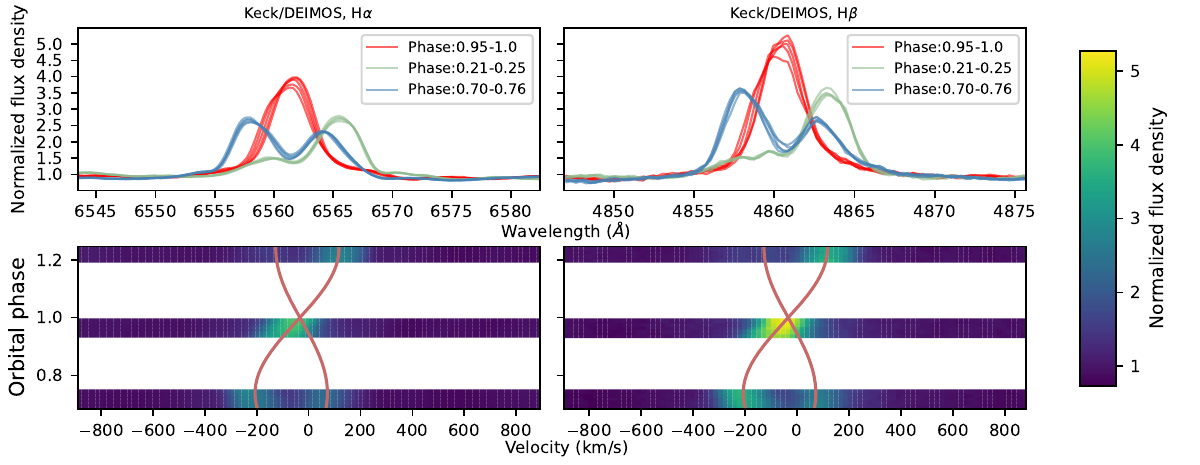}
    \caption{Dynamical spectra of J1553 from the Keck/DEIMOS observations. Upper panels: Line profiles of H$\alpha$ and H$\beta$ at all observation epochs. Lower panels: Dynamical spectra of the lines, phased with the orbital period of 4\,hr. Color scales indicate the continuum-normalized flux. The red sinusoidal curves represent the best-fit RV from Sect.~\ref{rv}.}
    \label{fig:spec}
\end{figure*}

In order to explore the physical properties of this system, on May 11 and 12, 2023 UTC we took a total of 14 phase-resolved spectra of J1553 using DEIMOS \citep{2003SPIE.4841.1657F} mounted on the Keck II 10\,m telescope on Maunakea with the 1200\,line\,${\rm mm}^{-1}$ grism . The wavelength coverage of the Keck spectra is 4700--6100\,\AA\ in the blue arm and 6100--7400\,\AA\ in the red arm. Each spectrum was taken with an exposure time of 120\,s, and a resolution of $R\approx 4000$. The DEIMOS spectra were reduced using the \texttt{LPIPE} reduction pipeline and \texttt{IRAF} routines. We performed wavelength calibration using the comparison-lamp spectra. The object spectra were then flux calibrated using spectrophotometric standard stars observed during the same night with the same instrumental configuration. We removed telluric lines from all spectra. Prominent TiO band absorption is observed in the spectra, typical  of M-dwarf stars. The spectra exhibit double-peaked H$\alpha$ and H$\beta$ emission lines, as shown in Fig.~\ref{fig:spec}; we can clearly see that the Balmer lines are modulated by orbital motion.

We used the full-spectrum fitting method developed by  \citet{2022MNRAS.513.4295K} to analyze the Keck spectra with a single-star assumption. We masked the H$\alpha$ and H$\beta$ lines and simultaneously fit the blue and red arms. The best-fit atmospheric parameters derived with minimal $\chi^2$ are $T_{\rm eff}=3215 \pm 188$\,K, log $g=4.63 \pm 0.08 $\,cgs, and [M/H] = $0.5 \pm 0.04$\,dex. This temperature matches that of an M4-type dwarf \citep{2013ApJS..208....9P}. The best-fit spectrum and the residuals are shown in Fig.~\ref{fig:spec_fitting}.

Next, we used the cross-correlation function (CCF; \citet{1979AJ.....84.1511T}) method implemented in the Python package \texttt{LASPEC} \citep{2020ApJS..246....9Z, 2021ApJS..256...14Z} to measure the RVs of J1553 and examine whether it harbors an invisible companion. The template used for the CCF method is the best-fit synthetic spectrum. We then calculated the CCF function at ${\rm RV}=v$  following Eq. 1 of \citet{2021ApJS..256...14Z}. The RV grid defined for the CCF varied from $-$800\,km\,s$^{-1}$ to 800\,km\,s$^{-1}$, with a step of 1\,km\,s$^{-1}$. The \texttt{Gaussian\_filter1d} function from \texttt{scipy} was used to smooth the CCF and compute its derivatives. We selected the RV range where the CCF exceeds the 75th percentile, and the RVs of the observed spectra were determined by the ascending zero point of the third derivative of the CCF in this range \citep{2017A&A...608A..95M, 2021ApJS..256...31L}. We estimated the uncertainties in RVs by Monte Carlo simulation. All RV measurements in this section were corrected for the barycentric rest frame.

We performed the CCF calculation in the 5000--6000\,\AA\ and 6900--7200\,\AA\  wavelength windows as a cross-check. The blue window (5000--6000\,\AA) contains several absorption lines, such as the Mg\,Ib triplet and Na\,I doublet, while the red window (6900--7200\,\AA) contains strong TiO absorption. From the third derivative of the CCF, we cannot see any signature of a second antiphased peak, so it should have only one visible component.

From the H$\alpha$ emission lines with clearly separated peaks, we 
estimated the locations of the peaks and measured the RVs of the antiphased component, according to the ascending zero point of the third derivative of the line profiles, with the same Gaussian filter function as mentioned above \citep{2007A&A...469..783S}. For the H$\beta$ lines at phases 0.21--0.25, the antiphased component does not show a clear profile, so we did not measure the RV of H$\beta$.

\section{Results} \label{sec:result}
\subsection{Radial velocity}
\label{rv}

\begin{figure}
 \centering
        \includegraphics[width=1\columnwidth]{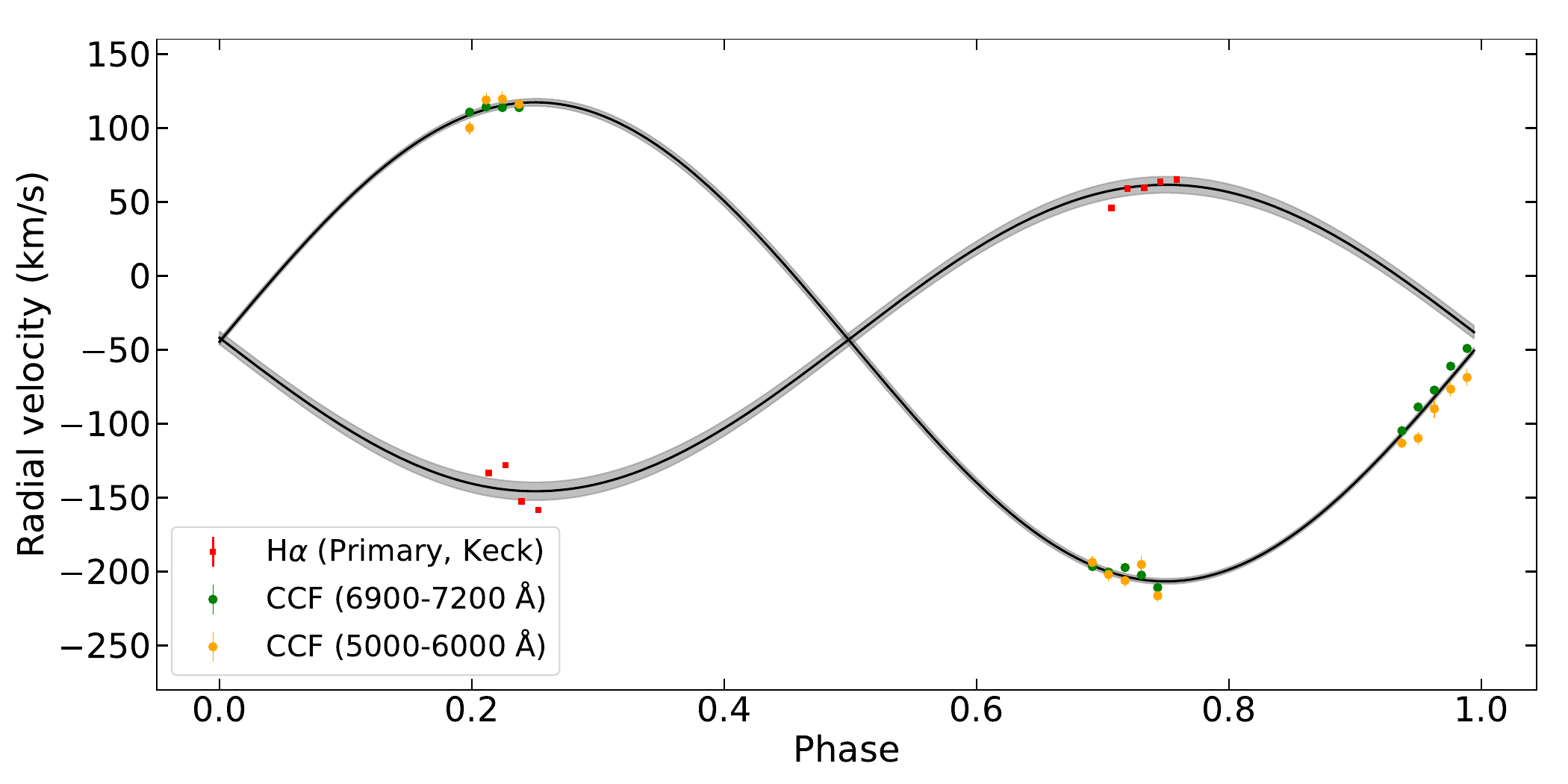}
    \caption{RV curves as a function of orbital phase. The black lines are the best-fit sinusoidal function for the Keck RV data from the CCF. The gray shaded region represents the 1$\sigma$ posterior spread. Orange and green dots represent RVs from the CCF method. Red squares represent RVs measured from H$\alpha$ lines.}
    \label{fig:rv}
\end{figure}

In principle, the RVs derived from the CCF method, which uses multiple spectral features, should be more accurate than those from single emission lines. We combined the CCF RVs from the blue arm and the red arm to calculate the RV semiamplitude of J1553B ($K_{\rm B}$). Considering that the orbital period (0.167\,d) is very short, we assumed that the orbit has been circularized \citep{1991A&A...248..485D, 2001IAUS..200...45M}. The CCF RV curve can be modeled as

\begin{equation}
\begin{aligned}
    {V_{{\rm RV}, j}}=K_{\rm B} \sin\big[\frac{2\pi}{P_{\rm orb}}\times (t_{{\rm obs}, j}- T_0)\big] - \gamma \, ,
 \label{eq:RV}
\end{aligned}
\end{equation}
where $K_{\rm B}$ is the RV semiamplitude, $t_{{\rm obs}, j}$ is the observed barycentric Julian date (BJD) of the $j$th spectrum, $T_0$ is the superior conjunction, and $\gamma$ is the systemic velocity of the binary system. We fitted the RV data with the Markov chain
Monte Carlo (MCMC) method. The final results give $K_{\rm B}=162.0\pm 1.5$\,km\,s$^{-1}$, $T_0=2459672.4664 \pm 0.0004$, and $\gamma=44.4\pm 1.4$\,km\,s$^{-1}$. From Fig.~\ref{fig:rv}, one can see that this sinusoidal function gives a reasonably good fit.

We performed RV curve fitting for the Keck RV data of H$\alpha$ from the antiphased component with the procedures described above, but using the result of $T_0$ as a prior. This yields a semiamplitude $K_{\rm A}=103.6 \pm 4.2$\,km\,s$^{-1}$ and a systemic velocity $\gamma_1 = 42.1 \pm 4.0$\,km\,s$^{-1}$. 

The mass function of the invisible star was calculated from
\begin{equation}
\begin{aligned}
    {f(M_{\rm A})}=\frac{M_{\rm A}^3\sin^3i}{(M_{\rm A}+M_{\rm B})^2} = \frac{K_{\rm B}^3P_{\rm orb}}{2\pi G}\, ,
 \label{eq:mass_function}
\end{aligned}
\end{equation}
which gives $f(M_{\rm A})= 0.074\pm 0.002 \, M_{\odot}$. This value is smaller than the value of $f(M_{\rm A})=0.083 \pm 0.002 \, M_{\odot}$ reported by  \citet{2024AJ....168..217D}. We note that their RVs were derived from the blue arm of LAMOST medium-resolution spectra. Our result should be more accurate because the Keck DEIMOS spectra we used have higher signal-to-noise ratios ($\sim100$) and wider wavelength coverage. 

\subsection{Light-curve modeling}

\label{subsec:solution}

We used the TESS light curve from Sector 51 in our light-curve modeling because it has less scatter and is closest to our Keck observations. The light curve was fitted simultaneously with the CCF RV curves using the \texttt{PHOEBE} 2.4 code \citep{2005ApJ...628..426P, 2016ApJS..227...29P,2020ApJS..250...34C} in a detached configuration, as indicated by the Doppler tomography in Sect.\ref{spec:tomography}. The primary star was set to be cool and small, with $R_1=3\times 10^{-6}\, R_{\odot}$ and $T_1=300$\,K. The gravity darkening coefficient of the visible star was set to be $g_2=0.32$ \citep{1967ZA.....65...89L}. We determined the logarithm bolometric and bandpass logarithmic limb-darkening coefficients by interpolating the grids from \citet{1993AJ....106.2096V} and \citet{2017A&A...600A..30C}. We integrated the MCMC sampler \texttt{EMCEE} \citep{2013PASP..125..306F} with the \texttt{PHOEBE} code.

\begin{figure}
        \includegraphics[width=1\columnwidth]{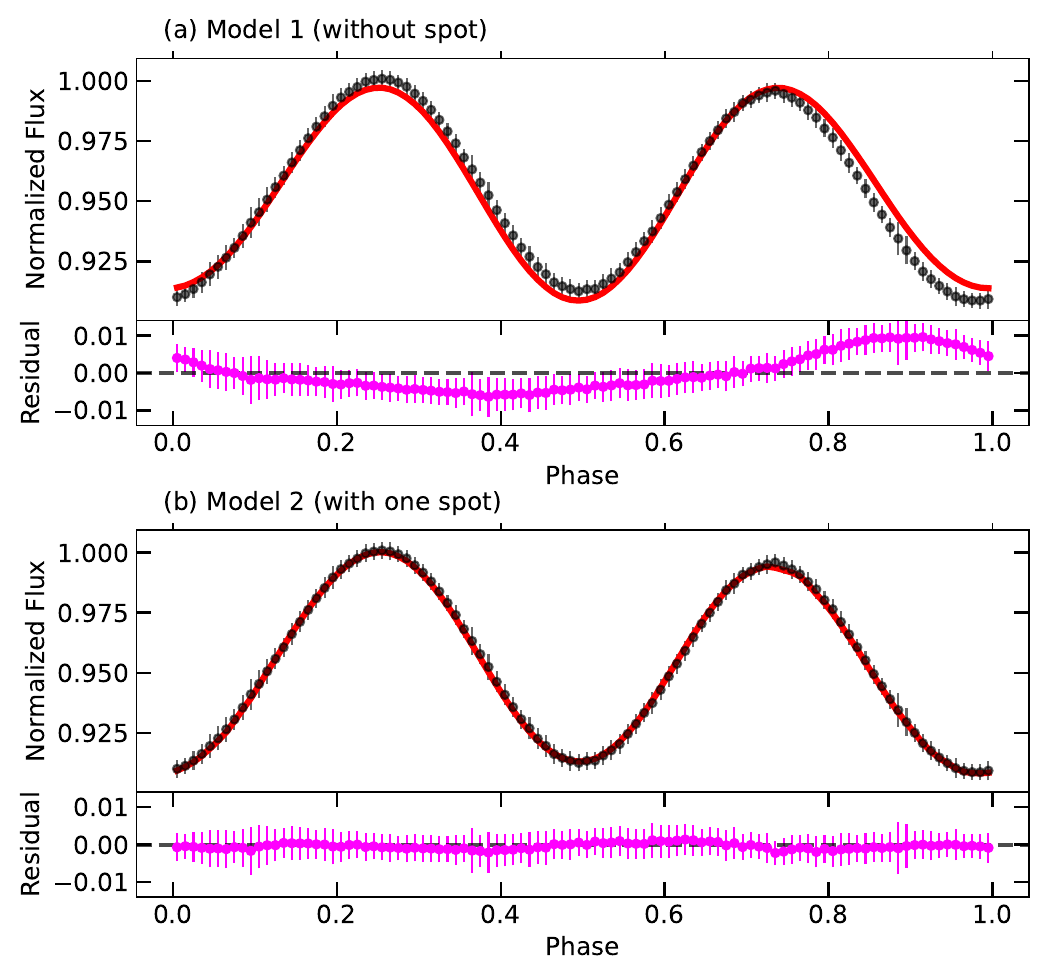}
    \caption{ Upper panel: Black points are the observed TESS light curve evenly divided into 100 phase bins. The error bars display the standard deviations of the bins. The red lines are theoretical light curves calculated by PHOEBE. Lower panel: Residuals between the observed and theoretical light curves. The error bars are the same as those in the upper panel.}
    \label{fig:lc_rv}
\end{figure}

The free parameters in the first model (Model 1) are as follows: mass of secondary $M_{\rm B}$;  superior conjunction $T_0$; semimajor axis $a$; orbital inclination $i$;  secondary radius $R_2$; and systemic radial velocity $\gamma$. To better constrain the parameters, we estimated the mass of J1553B, $M_{\rm B}$, according to the mass-luminosity relationship of \citet{2019ApJ...871...63M}. With the Two Micron All Sky Survey (2MASS) \textit{K}-band apparent magnitude $K_s=9.246 \pm 0.02$\,mag and the distance $d=36.833 \pm 0.017$\,pc \citep{2021AJ....161..147B}, we obtained $M_{\rm B}=0.37 \pm 0.02\,M_\odot$. Interstellar extinction was not considered, as $A_V$ given by GALExtin \citep{2021MNRAS.508.1788A} is nearly zero. We took this mass estimate as the prior distribution of $M_{\rm B}$, and the derived $T_0$ from the RV was also adopted as a prior distribution; the remaining parameters all had uniform prior distributions. We ran 40 walkers and 10,000 iterations for each walker during sampling. 
The mass of J1553A, the unseen WD, was calculated as $M_{\rm A}=(4\pi^2a^3)/(GP_{\rm orb}^2)-M_{\rm B}$. However, in Fig.~\ref{fig:lc_rv}, we can see that the light curve is not well modeled, showing periodic residuals.

In Model 2, we added a cool spot to account for variations in depth of two minima seen in the folded TMTS light curve relative to the TESS data. Because the area and the latitude of the spots are highly correlated with their temperature and radius \citep{2014MNRAS.442.2620Z}, we set the longitude, $\psi,$ and the angular radius, $r,$ of the spot as free parameters, and fixed the latitude, $\theta,$ and relative temperature, $T_{\rm spot}/T_{\rm B}$ , at $90^{\circ}$ and $0.9$, respectively. After including this spot, the fit improved and the periodic residuals were eliminated. For the upcoming calculation of the characteristics of J1553, we adopted the results of Model 2, which are shown in Table~\ref{tab1}.

\renewcommand{\arraystretch}{1.1}
\begin{table}
\setlength{\tabcolsep}{30pt}
\small
    \centering
        \caption{Parameters of light-curve solution. }
        \label{tab1}
 \begin{threeparttable}
        \begin{tabular}{lc} % four columns, alignment for each
                \hline
        \hline
        &  Model~2 (with one spot)\\
                Parameter &  Value \\
                \hline

  $M_{\rm A}$ ($M_{\sun}$) & $0.56\pm 0.09$\\
  $M_{\rm B}$ ($M_{\sun}$)  & $0.37\pm 0.02$\\
  $a$ ($R_{\sun}$) &  $1.25 \pm 0.04$\\
  $i$ (deg) &  $45.37^{+5.7}_{-4.9}$\\
  $T_0$ & $2459672.4661\pm 0.0001$\\
  $T_{\rm B}$ ($K$) & $3215^{*}$ \\
  $R_{\rm B}$ ($R_{\sun}$) & $0.403^{+0.014}_{-0.015}$ \\
  $\gamma$ (${\rm km/s}$) & $-43.3\pm 0.5$ \\
  
  \hline
  Spot 1 \\
  $\theta$ (deg) & $90^{*}$ \\
  $\psi$ (deg) & $214.45^{+4}_{-5}$ \\
  $r$ (deg) & $11.7^{+1.0}_{-0.8}$\\
  $T_{\rm spot}/T_{\rm B}$  & $0.9^*$\\

  \hline
  \hline
        \end{tabular}
 \begin{tablenotes}[normal, flushleft]
                \item 
                Note: The asterisk, $*,$ indicates that the value of the parameter is frozen in the fit.
        \end{tablenotes}
        \end{threeparttable}
\end{table}

\section{Discussion}\label{spec:discussion}

\subsection{The origin of Balmer emission lines}
\label{spec:tomography}

Figure~\ref{fig:spec} clearly shows that the Balmer emission lines exhibit a double-peaked profile. We can see that the stronger components in H$\alpha$ and H$\beta$ follow the sinusoidal RV curve of the absorption lines well, so they arise from the chromosphere of J1553B (the visible M dwarf). The weaker ones seem to trace the motion of J1553A (the invisible WD). In the cases of LTT~560 \citep{2007A&A...474..205T, 2011A&A...532A.129T} and SDSS 0138--0016 \citep{2012MNRAS.426.1950P}, these two binaries have a nearly Roche-filling secondary, and the WDs accrete through a stellar wind, so the systems exhibit double-peaked Balmer lines. \citet{2017MNRAS.470.4473P} also reported the measured gravitational redshifts for 16 eclipsing WD+MS binaries, including two sources with Balmer emission lines from the WD (SDSS J1021+1744 and SDSS J1028+0931). 
  
\begin{figure*}
\centering
        \includegraphics[width=1.0\columnwidth]{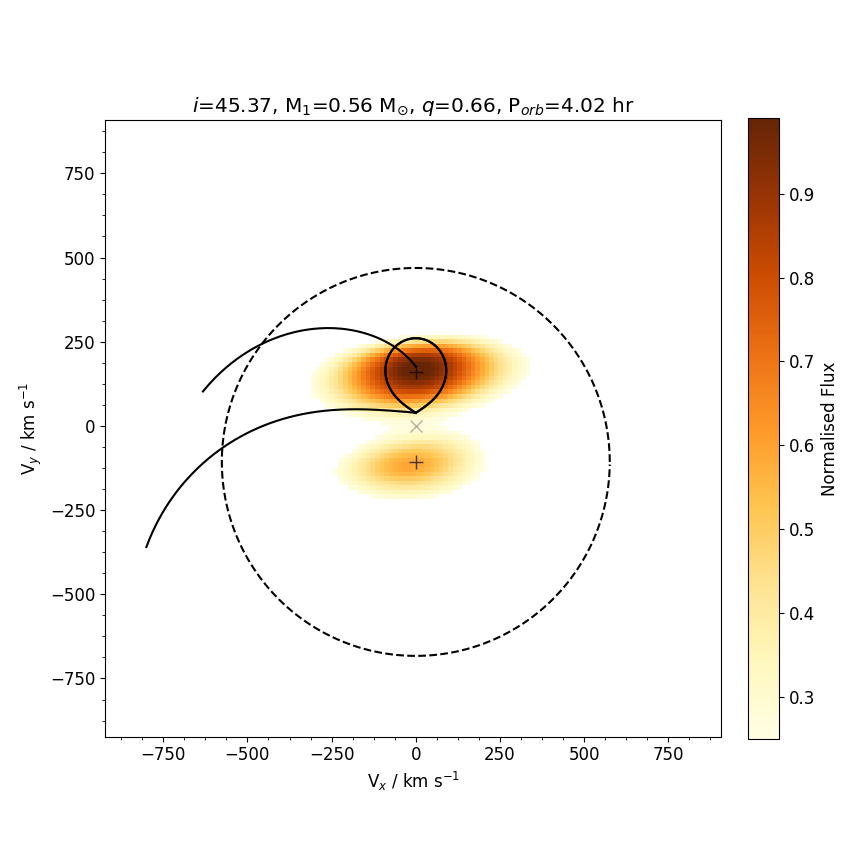}
    \includegraphics[width=1.0\columnwidth]{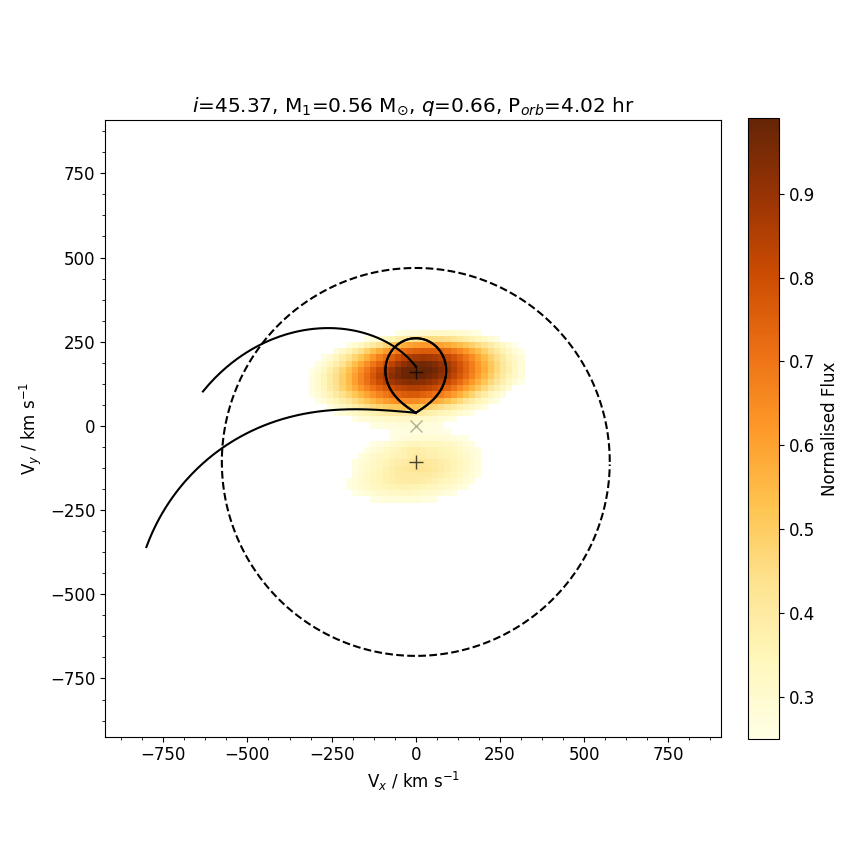}
    \caption{Doppler tomography of Balmer lines from the Keck observations. The positions of the primary and secondary are indicated by plus signs (+), and the center of mass of the system is labeled by a ``x'' mark. The solid lines represent the Roche lobe of the secondary and the stream trajectory. Color scales indicate the normalized flux. The dashed line represents the radius of the tidal limit of the accretion disk. Left: Tomography of H$\alpha$ lines. Right: Tomography of H$\beta$ lines.}
    \label{fig:tomography}
\end{figure*}

With $M_{\rm A}$ and $M_{\rm B}$ from Model 2, the mass ratio, $q,$ is $0.66\pm 0.11$, which is statistically consistent with $K_{\rm A}/K_{\rm B}=0.64 \pm 0.02$. In order to locate the Balmer emission regions, we used the Doppler tomography method to map the emission intensity of the time-series spectra in phase-velocity coordinates \citep{1998astro.ph..6141S}. This method is particularly powerful in  resolving accretion structures of interacting binaries via spectroscopic information \citep{2015A&A...579A..77K}. The Python package \texttt{pydoppler} \citep{2021ascl.soft06003H} was adopted in the analysis, which is a Python wrapper for Doppler tomography. In Fig. \ref{fig:tomography}, one can see there are two emission regions that overlap with the secondary and WD locations (``+'' signs in the plots). The tidal truncation radius of the accretion disk, which represents the maximum radius where the disk remains circular \citep{1995cvs..book.....W, 2023MNRAS.521.5846M, 2024MNRAS.531..422S}, is shown in Fig.\ref{fig:tomography}. There is no indication of the presence of an accretion disk. These facts further support the idea that antiphased emission lines that track the WD motion formed from its surface. This mass ratio can rule out the possibility that the primary is a MS star, as it would be more luminous than J1553B. 

We did not detect the difference between the systemic velocities of the CCF and H${\rm \alpha}$ RV curves in J1553 ($\gamma$ and $\gamma_1$ in Sect.\ref{rv}), which is the effect of gravitational redshift seen in the emission lines from the WD surface. The reason for this lack of detection is probably the limited spectral coverage and spectroscopic resolution; or perhaps the H$\alpha$ line originates from a region above the photosphere of the WD \citep{2011A&A...532A.129T}.

Given that J1553 is a detached binary system without an accretion-disk structure, the appearance of double-peaked Balmer emission lines in its spectra is unusual. The extra Balmer emission lines could be induced by stellar-wind accretion as depicted above. Detecting stellar winds of late-type MS stars is difficult \citep{2024ApJ...976...65R, 2024NatAs...8..596K}, but some progress has been made. For instance, \citet{2021ApJ...915...37W} successfully measured the wind mass-loss rate for 9 M-dwarf stars; most of them have $\dot{M}\lesssim 1 \dot{M}_{\odot}$ ($\dot{M}_{\odot}=2\times 10^{-14}\,M_{\odot}\,{\rm yr}^{-1}$), which increases the constraints on M-dwarf winds. 

To better quantify these emission lines, we applied a two-Gaussian function to fit the H$\alpha$ and H$\beta$ lines of the Keck spectra covering phases 0.25 and 0.75. The ratio of H$\alpha$ and the bolometric luminosity can be calculated as $L_{\rm H\alpha}/L_{\rm bol}= \chi \times {\rm EW}$ \citep{2014ApJ...795..161D}, where EW is the mean equivalent width of the H$\alpha$ line from J1553A. Here we adopt a $\chi=3.5926\times 10^{-5}$ for an M4-type star (see Table 7 of \citealt{2014ApJ...795..161D}).
Using the Stefan-Boltzmann law, the bolometric luminosity of J1553B is $L_{\rm bol, B}= (T_{\rm B}/T_{\sun})^4 \, (R_{\rm B}/R_{\sun})^2 \, L_{\sun} = (5.9\pm 1.5) \times 10^{31}$\,erg\,s$^{-1}$. We measured ${\rm EW}_{\rm H\alpha, A}=4.0 \pm 0.2$\,\AA, so the $ {\rm H\alpha}$ luminosity is $L_{\rm H\alpha, A}= \chi\, {\rm EW}_{\rm H\alpha, A}\, L_{\rm bol,B}= ({8\pm 2})\times10^{27}$\,erg\,s$^{-1}$. We also calculated the H$\beta$ luminosity of J1553A; with $\chi_{\rm H\beta}=0.167\times 10^{-4}$ \citep{2008PASP..120.1161W} and ${\rm EW}_{\rm H\beta, A}=4.8 \pm 0.2$ \AA, $L_{\rm H\beta, A}$ is derived to be $L_{\rm H\beta, A}= (5\pm 1)\times10^{27}$\,erg\,s$^{-1}$. Thus, the emission lines originating from the WD have a luminosity of at least $\sim 10^{28}$\,erg\,s$^{-1}$, which is higher than the H${\alpha}$ luminosity reported in LTT~560. Assuming that all the gravitational energy of the accreting material is released as radiation,  for a WD with $M=0.56\, M_{\odot}$ and $R=0.02\, R_{\odot}$, the  luminosity of extra Balmer emission requires an accretion rate of $\dot{M}= 3\times 10^{-15}\,M_{\odot}\,{\rm yr}^{-1}$. Thus, even with a high efficiency to accrete material, the wind mass-loss rate should satisfy $\dot{M}_{\rm wind}\gtrsim 0.15\, \dot{M}_{\odot}$. In addition, this system might allow for detailed investigation of heating processes occurring near the surface of the WD during accretion \citep{2004ApJ...600..390T}. 

\subsection{The origin and the outcome of J1553}
\label{evlove}

The progenitor origin of J1553 remains uncertain. However, we could speculate that because of the short orbital period of J1553, one possible progenitor could be a binary system involving an M dwarf of $0.37\,{M}_{\odot}$ and a larger mass MS star. When the MS star evolves to an asymptotic giant branch (AGB) star, in which the CO core increases to a mass of $\sim0.56\,{M}_{\odot}$, the system enters the CE phase, after which the envelope is ejected, and the system now observed is formed. 

During the CE phase, the reduced orbital energy is used to eject the CE. The standard energy budget formula for the CE phase is (e.g., \citealt{1984ApJ...277..355W})

\begin{equation}
\label{eq:alpha}
\begin{split}
    {\alpha}_{\rm {CE}}\big(\frac{G{M}_{\rm {core}}{M}_{\rm {dwarf}}}{2{a}_{\rm f}}-\frac{G({M}_{\rm {core}}+{M}_{\rm {env}}){M}_{\rm {dwarf}}}{2{a}_{\rm i}}\big)\\= \frac{G{M}_{\rm d,i}{M}_{\rm {env}}}{{\lambda}{R}_{\rm {d,i}}}\, ,
\end{split}
\end{equation}

where the left side is the release of orbital energy, the right side is the binding energy of the envelope, ${M}_{\rm {core}}$ and ${M}_{\rm {env}}$ respectively represent the core mass and envelope mass of the MS, ${M}_{\rm {dwarf}}$ is the mass of the M dwarf, ${M}_{\rm {d,i}}$ and ${R}_{\rm {d,i}}$ are respectively the initial mass and radius of the MS, and ${a}_{\rm f}$ and ${a}_{\rm i}$ are respectively the final and initial binary separations of the system. According to \citet{1983ApJ...268..368E}, ${a}_{\rm i}$ can be defined as
\begin{equation}
\label{eq:separation}
    {a}_{\rm i}={R}_{\rm {RL}}\frac{0.6{q}^{2/3}+\ln(1+{q}^{1/3})}{0.49{q}^{2/3}}\, ,
\end{equation}
where ${R}_{\rm {RL}}={R}_{\rm {d,i}}$ and $q={M}_{\rm {d,i}}/{M}_{\rm {dwarf}}$. For ${a}_{\rm f}$, because we know the system parameters, we can easily estimate ${a}_{\rm f}\approx1.248\,{R}_{\odot}$.

To investigate the origin of J1553, we evolved MS stars with different masses toward their AGB phases until the CO cores increased to a mass of $0.56\,{M}_{\odot}$ (for J1553A), with an attempt to examine whether the CE ejection efficiency, ${\alpha}_{\rm {CE}}$ , is in a reasonable range (i.e., 0.0--1.5). In the subsequent analysis, we adopt the following assumptions: (1) these AGB stars fill their Roche lobe when their cores reach a mass of $0.56\,{M}_{\odot}$ (i.e., ${R}={R}_{\rm RL}$), and they enter the CE phases with the companion (i.e., J1553B, the $0.37\,{M}_{\odot}$ MS star); (2) considering the function of Eq. \ref{eq:alpha}, after the CE ejection, the system should have the same parameters as observed (i.e., ${a}_{\rm f}\approx1.248\,{R}_{\odot}$); and (3) the stellar structure parameter, ${\lambda,}$ is assumed to be 1.0 for a fixed mass of the MS star.

We used the stellar evolution code Modules for Experiments in Stellar Astrophysics ($\tt{MESA}$; \citealt{2011ApJS..192....3P,2019ApJS..243...10P}) to explore the progenitor system of J1553 and found that it is consistent with an initial mass of 2.0-4.0\,${M}_{\odot}$ considering the reasonable range required for CE efficiency, ${\alpha}_{\rm {CE}}$.
For a mass of 2.0\,${M}_{\odot}$, the initial separation of ${a}_{\rm i}$ is found to be $\sim 342\,{R}_{\odot}$, with an orbital period of ${P}_{\rm i}\approx476\,{\rm d}$) and ${\alpha}_{\rm CE}\approx0.192$. For a $3.0\,{M}_{\odot}$ MS star, ${a}_{\rm i} \approx 289\,{R}_{\odot}$, with ${P}_{\rm i}\approx310\,{\rm d}$ and ${\alpha}_{\rm CE}\approx0.54$. For a $4.0\,{M}_{\odot}$ MS star, ${a}_{\rm i}$ is estimated to be $\sim236~{R}_{\odot}$ (${P}_{\rm i}\approx201\,{\rm d}$), and ${\alpha}_{\rm CE}\approx1.2$. In this case, we suggest that the origin of J1553 is a binary system involving a 2.0--4.0\,${M}_{\odot}$ MS star and a $0.37\,{M}_{\odot}$ M-dwarf star; when the MS star evolves to the AGB, they enter the CE phase.

The radius of the Roche lobe of J1553B is  $\sim 0.429\,{R}_{\odot}$, so it is a star that almost fills its Roche lobe. We simulated the further evolution of J1553; the WD was treated as a point mass, and we evolved a $0.37\,{M}_{\odot}$ MS star with solar metallicity until its radius reached $0.403\,{R}_{\odot}$. Then we put the system in a ${P}=0.16756456$\,d orbit. During the simulation, we considered the loss of angular momentum via mass loss, magnetic braking, and gravitational wave radiation. For the magnetic braking mechanism,  we adopted the widely used RVJ law \citep{1983ApJ...275..713R}. Since the mass accretion onto the WD may lead to a CV, we considered the mass-accretion efficiency of the WD based on \citet{2004ApJ...613L.129K}; that is, if the mass accretion rate is higher than the critical mass accretion rate ($\dot{M}_{\rm {cr}}$),
\begin{equation}
    \dot{M}_{\rm {cr}}=5.3\times{10}^{-7}\frac{(1.7-X)}{X}({M}_{\rm {WD}}/{M}_{\odot}-0.4)\,{M}_{\odot}\,{\rm {yr}}^{-1},
\end{equation}
where ${X}$ is the hydrogen mass fraction and ${M}_{\rm {WD}}$ is the mass of the accreting WD, the unprocessed material is assumed to be lost from the system as an optically thick wind. If the mass accretion rate is less than $\dot{M}_{\rm {cr}}$ but higher than $(1/2)\,\dot{M}_{\rm {cr}}$, the H-shell burning is steady and no mass is lost from the system. If the mass accretion rate is lower than $(1/2)\,\dot{M}_{\rm {cr}}$ but higher than $(1/8)\,\dot{M}_{\rm {cr}}$, a very weak H-shell flash could be triggered, but no mass is lost from the system. If the mass accretion rate is lower than $(1/8)\,\dot{M}_{\rm {cr}}$, the H-shell flash is too strong to accumulate material on the WD surface. We define the mass growth rate of the He layer ($\dot{M}_{\rm He}$) below the H-shell burning as 
\begin{equation}
    \dot{M}_{\rm {He}}={\eta}_{\rm H}\dot{M}_{\rm {acc}}\, ,
\end{equation}
where ${\eta}_{\rm H}$ is the mass accumulation efficiency of H-shell burning and $\dot{M}_{\rm {acc}}$ is the mass accretion rate of the WD. When the mass of the He layer reaches a certain value, helium is assumed to be ignited. If He-shell flashes occur, a part of the envelope mass is assumed to be blown off. In this case, the WD mass growth rate is derived from the fitting formula provided by \citet{2017A&A...604A..31W}. Hence, the mass growth rate of the WD ($\dot{M}_{\rm CO}$) is 
\begin{equation}
    \dot{M}_{\rm {CO}}={\eta}_{\rm H}{\eta}_{\rm {He}}\dot{M}_{\rm {acc}}\, .
\end{equation}

Based on MESA simulations, we found that J1553B will fill its Roche lobe after  $\sim 3.3\times{10}^{6}$\,yr. At the onset of mass transfer, the transfer rate is $\dot{M} \approx{10}^{-10}\,{M}_{\odot}\,{\rm yr}^{-1}$.  
Subsequently, this rate rapidly increases to $\dot{M}\approx{10}^{-8}\,{M}_{\odot}\,{\rm yr^{-1}}$ owing to the loss of orbital angular momentum via magnetic braking and gravitational wave radiation. The system may appear as a supersoft X-ray binary when the accreted H-rich material can burn steadily on the surface of the WD. This stable burning phase of H can last about $1.05\times{10}^{6}$\,yr. During this phase, the orbital period will lengthen as a result of the increase in WD mass \citep{2020AJ....159..189L, 2025AJ....169..139X}. 
After the short-lived supersoft X-ray stage, the mass-transfer rate drops rapidly to $\sim {10}^{-9}\,{M}_{\odot}\,{\rm yr^{-1}}$, and the system will appear as a CV. In this phase, the WD accretes the H-rich material at a low mass accretion rate (lower than $(1/8)\,\dot{M}_{\rm {cr}}$), with no mass accumulated on the WD. At $P_{\rm orb}\approx 3$\,hr, the M dwarf will become fully convective as the mass decreases to 0.2\,$M_{\odot}$ \citep{2006MNRAS.373..484K}. Then the magnetic braking stops and J1553 will evolve following the standard evolutionary track of a CV \citep{2011ApJS..194...28K}.  

\section{Conclusion} \label{sec:conclusion}

We discovered a PCEB, J1553, from the minute-cadence survey of TMTS. This binary system contains an invisible WD and an M dwarf, orbiting with a period $P_{\rm orb}=0.16756456$\,d. J1553 is 
a single-lined spectroscopic binary according to the single-peaked CCF curves. We estimated the detailed parameters of J1553 by applying a simultaneous fit to the light curve and the CCF RV data. The derived mass of J1553A is $M_{\rm A}=0.56\pm 0.09\,M_{\odot}$, while the mass and radius of J1553B are $M_{\rm B}=0.37\pm 0.02\, M_{\odot}$ and $R_{\rm B}=0.403^{+0.014}_{-0.015}\,R_{\odot}$, respectively. 

Except for the Balmer emission lines from J1553B, the spectra exhibit antiphased extra emission lines, likely owing to their formation on the WD surface. The total luminosity of the H$\alpha$ and H$\beta$ lines of J1553A can reach $\sim 10^{28}$\,erg\,s$^{-1}$, relatively high compared with LTT~560. We explored the underlying physical process of the hydrogen radiation. 

Based on the system lacking a disk at the present stage, we suggest this binary is a detached system, and the secondary is more massive than the previous estimate of 0.29\,$M_{\odot}$ from the light-curve fitting in such a configuration \citep{2024AJ....168..217D}. This difference in mass estimate has a critical influence on the magnetic braking effect \citep{2010A&A...513L...7S}, and consequently on the evolutionary path.

With a MESA simulation, we inferred that J1553 originated from a binary with a 2--4\,$M_{\odot}$ star and a $0.37\,M_{\odot}$ M dwarf by constructing the CE phase using the standard energy budget formula and assuming the stellar structure parameter $\lambda=1$.
In its future evolution, J1553 is predicted to become a supersoft X-ray binary with a lifetime of about $10^{6}$\,yr, and then it will appear as a CV.

J1553 is an excellent system for studying wind accretion and binary evolution theory. Moreover, it is a valuable source that has the potential to provide precise measurements of both mass and radius of the dark WD with the help of higher-resolution spectra in the future.  

\begin{acknowledgements}
      This work is supported by the National Science Foundation of China (NSFC grants 12288102 and 12033003), the Ma Huateng Foundation, and the Tencent Xplorer Prize. J.L. is supported by the National Natural Science Foundation of China (NSFC; grant  12403038), the Fundamental Research Funds for the Central Universities (grant  WK2030000089), and the Cyrus Chun Ying Tang Foundations. C. Wu is supported by the National Science Foundation of China (NSFC grant 12473032), the National Key R\&D Program of China (No. 2021YFA1600404), the Yunnan Revitalization Talent Support Program—Young Talent project, and the International Centre of Supernovae, Yunnan Key Laboratory (No. 202302AN360001). A.V.F.'s team received support from the Christopher R. Redlich Fund, Gary and Cynthia Bengier, Clark and Sharon Winslow, Alan Eustace (W.Z. is a Bengier-Winslow-Eustace Specialist in Astronomy), William Draper, Timothy and Melissa Draper, Briggs and Kathleen Wood, Sanford Robertson (T.G.B. is a Draper-Wood-Robertson Specialist in Astronomy), and many other donors. C. Liu is supported by the Beijing Natural Science Foundation (No. 1242016), Talents Program (24CE-YS-08), and the Popular Science Project (24CD012) of the Beijing Academy of Science and Technology. H.G. acknowledges support from the National Key R\&D Program of China (grant 2021YFA1600403), NSFC (grant 12173081), and the key research program of frontier sciences, CAS (No. ZDBS-LY-7005). J. Li is supported by the National Science Foundation of China (NSFC grant 12403040).

This paper includes data collected with the TESS mission, obtained from the MAST data archive at the Space Telescope Science Institute (STScI). Funding for the TESS mission is provided by the NASA Explorer Program. STScI is operated by the Association of Universities for Research in Astronomy, Inc., under NASA contract NAS 5–26555.

This work made use of observations obtained with the Samuel Oschin 48-inch telescope and the 60-inch telescope at Palomar Observatory as part of the Zwicky Transient Facility project. ZTF is supported by the U.S. National Science Foundation (NSF) under grants AST-1440341 and AST-2034437 and~collaboration including current partners Caltech, IPAC, the~Weizmann Institute for Science, the~Oskar Klein Center at Stockholm University, the~University of Maryland, Deutsches Elektronen-Synchrotron and Humboldt University, the~TANGO Consortium of Taiwan, the~University of Wisconsin at Milwaukee, Trinity College Dublin, Lawrence Livermore National Laboratories, IN2P3,
University of Warwick, Ruhr University Bochum, and~Northwestern University, and~former partners the University of Washington,
Los Alamos National Laboratories, and~Lawrence Berkeley National
Laboratories. Operations are conducted by COO, IPAC, and~UW. utions participating in the Gaia Multilateral Agreement.
\end{acknowledgements}

\bibliographystyle{aa} 
\bibliography{aa53732-25.bib} 

\begin{appendix} 
\section{Spectroscopic data}

\begin{figure}[h]
 \centering
        \includegraphics[width=1\columnwidth]{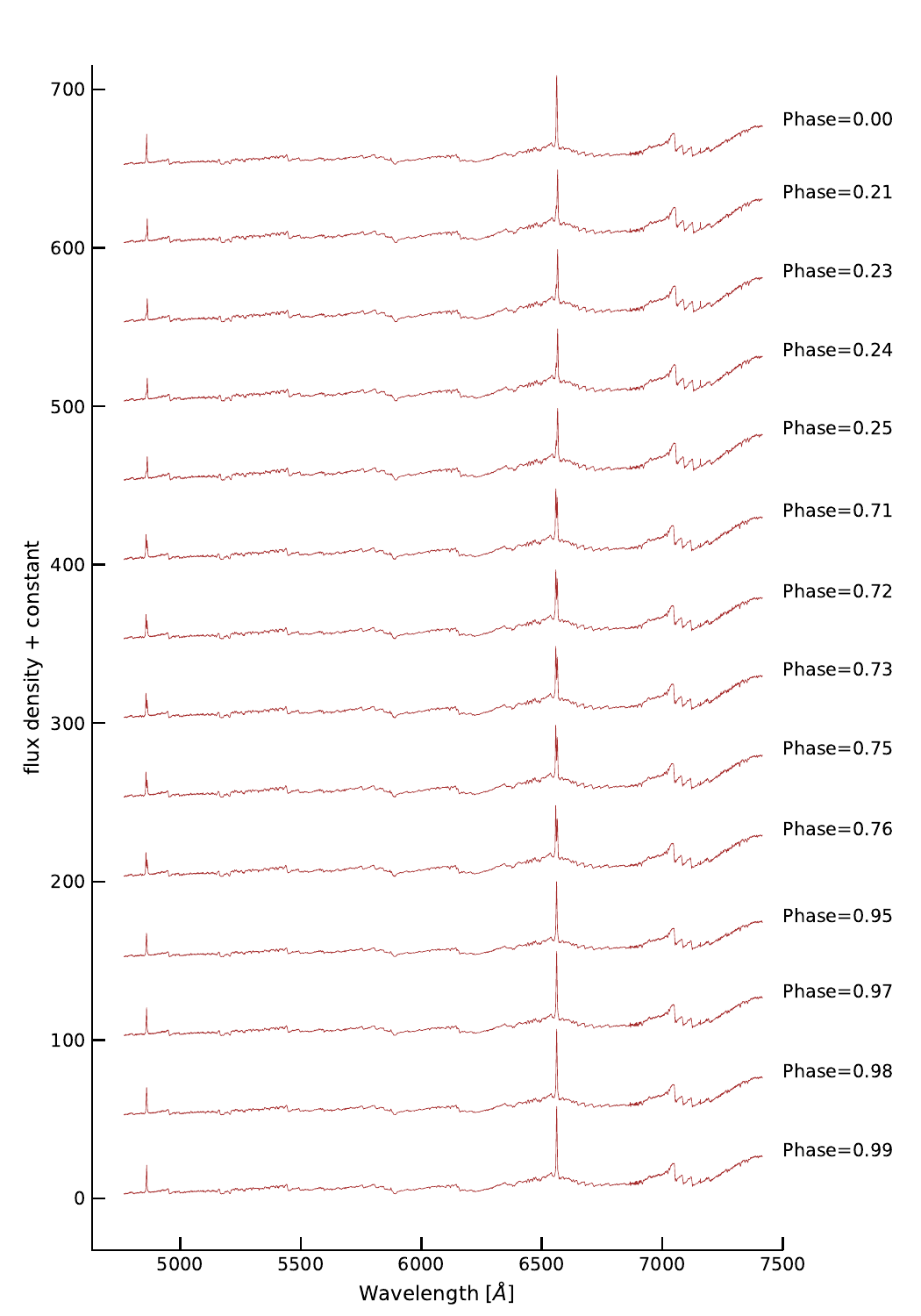}
    \caption{The observed time-series Keck spectra. The corresponding phases are labeled in the right of each spectrum.}
    \label{fig:spec_plot}
\end{figure}

\begin{figure}[h]
 \centering
        \includegraphics[width=1\columnwidth]{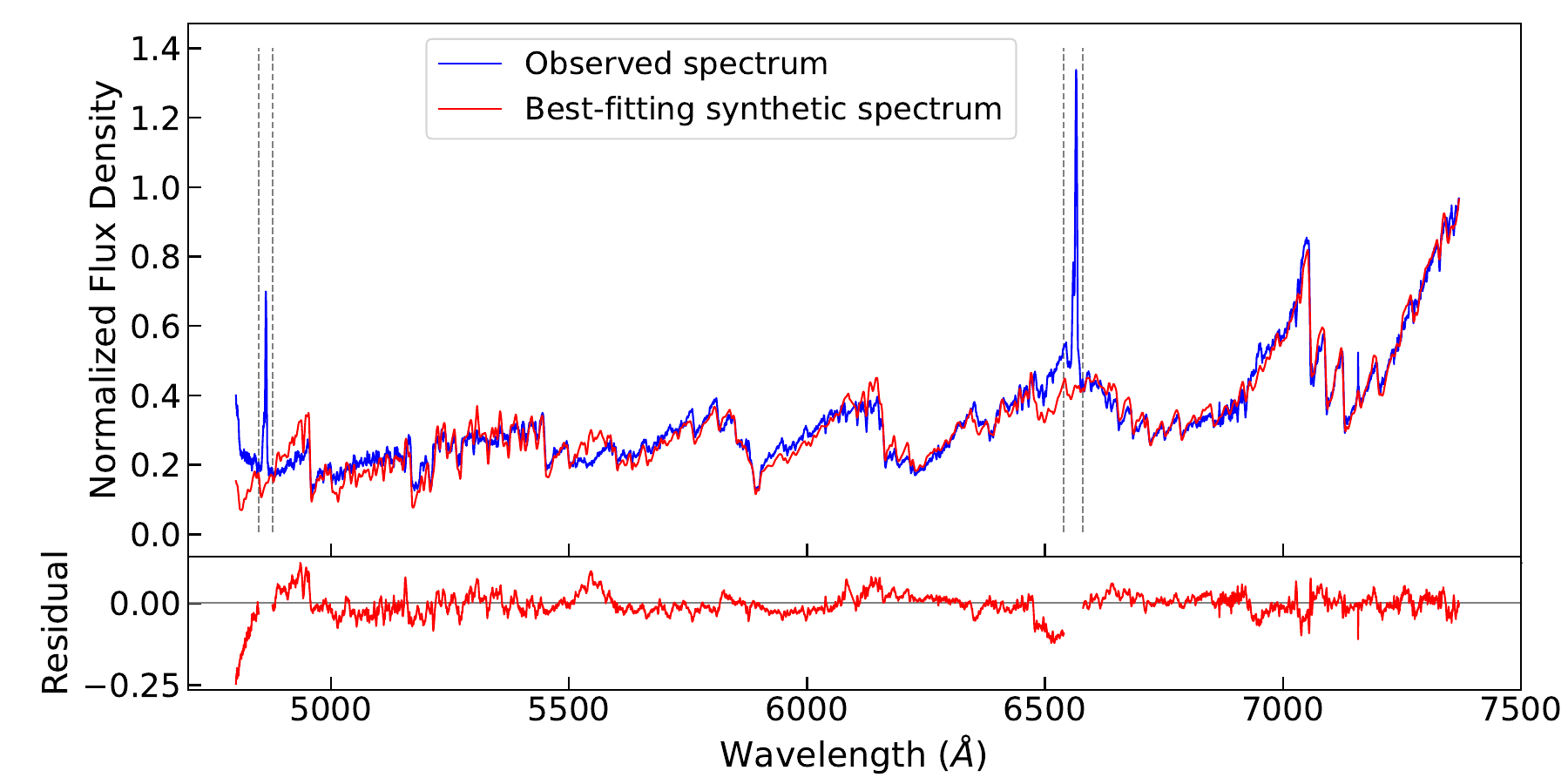}
    \caption{Upper panel: the observed Keck spectrum (solid blue line) and the best-fit synthetic spectrum (solid red line). The dashed gray regions indicate the masked emission lines. Lower panel: the residual between the Keck spectrum and the synthetic spectrum. }
    \label{fig:spec_fitting}
\end{figure}

The time-series Keck spectra are displayed in Fig.\ref{fig:spec_plot}. The best-fit synthetic spectrum, derived using the method described in Sect.\ref{sec:floats}, is illustrated in Fig.\ref{fig:spec_fitting}.  The single-star template shows a reasonable agreement with the observed spectrum, except for the emission lines.

\end{appendix}

\end{document}